\documentclass[nocopyright]{cimento}

\usepackage{graphicx}
\usepackage{amsmath}
\usepackage{amssymb}
\usepackage{latexsym}
\usepackage{tabularx} % per tabelle
\usepackage{multirow}

\title{Einstein Telescope: Ferromagnetic Shielding for Magnetic Noise Mitigation}
\author{Federico Armato\from{ins:x}\from{ins:y}\thanks{federico.armato@ge.infn.it}, Barbara Garaventa\from{ins:x}, Andrea Chincarini\from{ins:x}}
\instlist{\inst{ins:x} INFN, Sezione di Genova - Genova, Italy
  \inst{ins:y} Dipartimento di Fisica, Universit\`a di Genova - Genova, Italy}

\makeatletter
\renewenvironment{table}[1][\fps@table]
 {%
  \begingroup\edef\x{\endgroup\noexpand\@float{table}[#1]}\x
  \small\@intable@true
  \def\tnote##1{\par$({##1})$}%
 }
 {\@intable@false\end@float}
\makeatother

\begin{document}

\maketitle

\begin{abstract}
The Einstein Telescope is the next-generation gravitational wave interferometer which, compared to current detectors, will enable the observation of gravitational signals at lower frequencies with a sensitivity improved by approximately two orders of magnitude. Achieving such exceptional sensitivity requires minimizing all sources of noise. In the low-frequency regime, magnetic noise is one of the dominant. This article examines the effectiveness and limitations of a passive mitigation technique: ferromagnetic shielding.
\end{abstract}

\section{Introduction}\label{Introduction}
\begin{figure}[h!]
    \centering
    \includegraphics[width=0.5\linewidth]{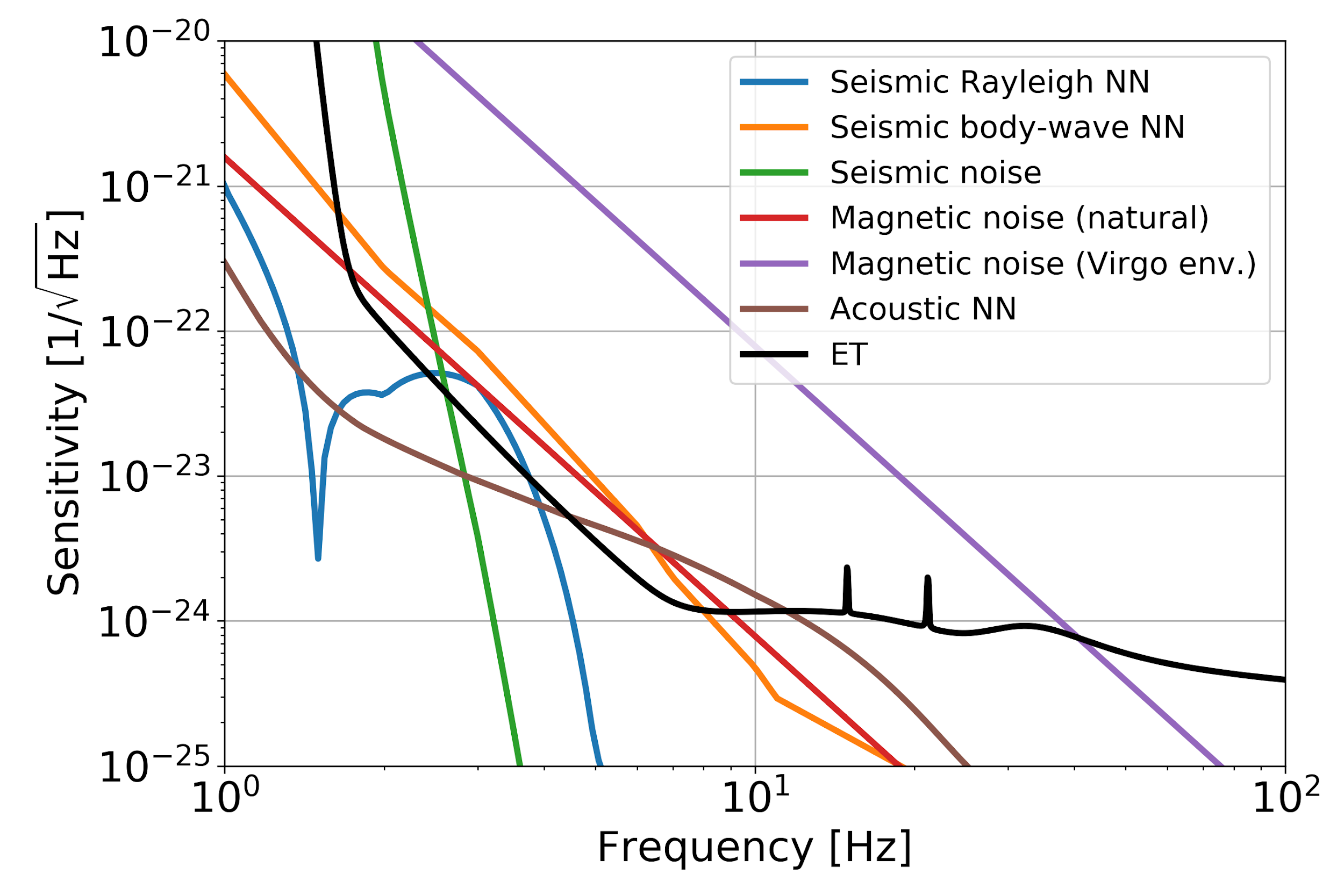}
    \caption{\footnotesize Example of an ET environmental noise budget together with the ET sensitivity curve~\cite{Virgo_ET}.}
    \label{fig:sensitivity_curve}
\end{figure}

In the future Einstein Telescope (ET) gravitational wave interferometer, magnetic noise - if left unaddressed - is expected to be one of the dominant noise sources at low frequencies~\cite{SiteSelection}.\\[0.1cm] 
Indeed, to reach the design sensitivity of the ET, this noise must be reduced by at least a factor of 3 for naturally occurring sources, and by at least a factor of 100 - taking Virgo as a reference - for self-inflicted noise (Fig.~\ref{fig:sensitivity_curve}).\\[0.1cm]
Magnetic noise mitigation is implemented through dedicated techniques, which can be categorized as either active, such as adaptive current compensation systems, or passive, including eddy currents and ferromagnetic shielding.\\[0.1cm]
Among passive methods, eddy currents prove to be ineffective at low frequencies, yielding an extremely low shielding factor. In contrast, ferromagnetic shielding - even with a single layer - provides a markedly more effective attenuation~\cite{ArmatoMagneticNoise}.\\[0.1cm]
%To provide a clear overview of the study, this article is structured as follows. Section \ref{Magnetic Noise} introduces the concept of magnetic noise and its impact on gravitational wave detectors. Section \ref{Ferromagnetic Shielding} presents ferromagnetic shielding as a potential mitigation strategy. This approach is then applied to a test mass tower in Section \ref{Test Mass Towers}, with the results presented in Section \ref{Results}. Finally, Section \ref{Conclusion} summarizes the main findings and discusses possible directions for future work.

%%%%%%%%%%%%%%%%%%%%%%%%%%%%%%%%%%%%%%%%%%%%

\section{Magnetic Noise}\label{Magnetic Noise}
The term magnetic noise can be somewhat misleading. Unlike other noise sources - such as quantum or thermal noise - which are intrinsic to the interferometer, magnetic noise originates from the interaction between external magnetic fields and specific components of the interferometer. Without these coupling mechanisms, magnetic noise would be entirely absent.\\[0.1cm]
As such, the study of magnetic noise requires a comprehensive analysis of both the sources of magnetic fields and the interferometer components that interact with them, along with a detailed understanding of the coupling mechanisms involved.\\[0.1cm]
Consequently, magnetic noise mitigation relies on a triple strategy:
\begin{itemize}
    \item Reduction of the magnetic field at the source (typically achievable when the source is anthropogenic);
    \item Attenuation of the magnetic field in proximity to interferometer components that are sensitive to external fields.
    \item Reduction of the coupling factor.
\end{itemize}

%%%%%%%%%%%%%%%%%%%%%%%%%%%%%%%%%%%%%%%%%%%%

\section{Ferromagnetic Shielding}\label{Ferromagnetic Shielding}
Ferromagnetic shielding utilizes materials with high magnetic permeability to redirect magnetic field lines. This technique is used both to reduce a device’s magnetic emissions - by using a ferromagnetic coating to confine part of its magnetic field to the surrounding area - and to protect sensitive regions of an interferometer - minimizing the influence of external magnetic fields.\\[0.1cm]
The materials typically used are Ni-Fe alloy systems, such as mu-metal and supermalloy, known for their extremely high magnetic permeability and exceptionally narrow hysteresis loop - key characteristics that prevent residual magnetic fields ({\footnotesize T}{\scriptsize ABLE}~{\footnotesize\ref{table:Fe-Ni alloy}}).\\
\begin{table}[h!]
    \centering
    \caption{\footnotesize Comparison between soft iron and two different Fe-Ni alloys. $\mu_i$ and $\mu_{max}$ are the initial and maximum relative permeability, respectively, reached by the materials, and $H_c$ is the coercivity field~\cite{Fe-Ni_Alloy}.}
    \label{table:Fe-Ni alloy}
    \begin{tabularx}{\textwidth}{p{0.28\textwidth}p{0.2\textwidth}p{0.13\textwidth}p{0.13\textwidth}p{0.13\textwidth}}
    \hline
    \rule{0pt}{15pt} \textbf{Material} & \textbf{Name} & \boldmath{$\mu_i$} & \boldmath{$\mu_{max}$} & \boldmath{$H_c$} \\ [0.5ex] 
    \hline\hline
    \rule{0pt}{12pt} $Fe$ & Soft iron & $300$ & $5\,000$ & $70\,A/m$\\
    \rule{0pt}{12pt} $Ni_{77}Fe_{16.5}Cu_{5}Cr_{1.5}$ & mu-metal &  $20\,000$ & $100\,000$ & $4\,A/m$\\
    \rule{0pt}{12pt} $Ni_{180}Fe_{15}Mo_{5}$ & supermalloy &  $100\,000$ & $300\,000$ & $0.5\, A/m$\\[0.1cm]
 \hline
    \end{tabularx}
\end{table}\\
The magnetic response of these materials depends on both the intensity and frequency of the external magnetic field. As reported in \cite{Schumann_Virgo}, \cite{Magnetic_Noise_Virgo} and \cite{ET-0069A-22}, the magnetic fields measured at Virgo and at Sos Enattos - one of the proposed sites for the Einstein Telescope - are below the $nT$ level. However, for such low field strengths, direct measurements in the literature are scarce. For instance, \cite{Arpaia} reports measurements of the magnetic permeability of an alloy composed of 80\% Ni, 5\% Mo, 15\% Fe, 0.3–0.5\% Mn, and 0.1–0.4\% Si down to a few $nT$, while \cite{mu-metal} presents measurements for an alloy of 80\% Ni, 15\% Fe, 4.5\% Mo, 0.4\% Mn, and 0.1\% Si down to the $\mu T$ range.

%%%%%%%%%%%%%%%%%%%%%%%%%%%%%%%%%%%%%%%%%%%%

\section{Test Mass Towers}\label{Test Mass Towers}
\begin{figure}[h!]
\centering
\begin{minipage}[c]{0.44\textwidth}
\centering
    \includegraphics[width=0.42\textwidth]{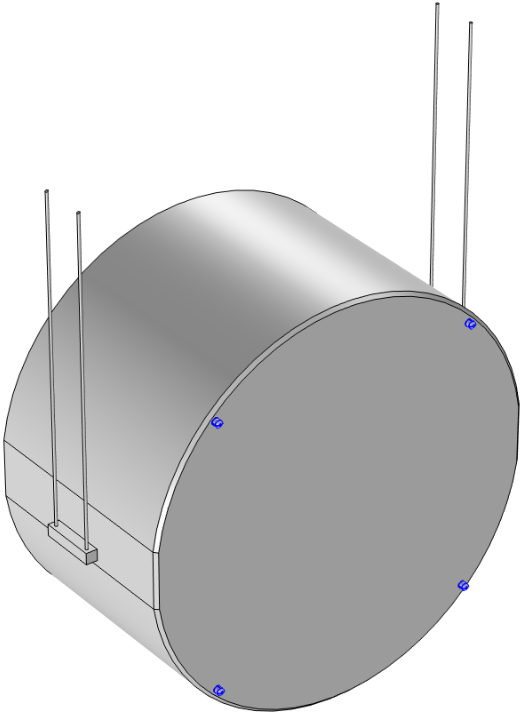}
    \caption{\footnotesize CAD drawing of a TM and the four magnets glued on it (in blue).}
    \label{fig:TM_magnets}
\end{minipage}
\hfill
\begin{minipage}[c]{0.44\textwidth}
\centering
    \includegraphics[width=0.9\textwidth]{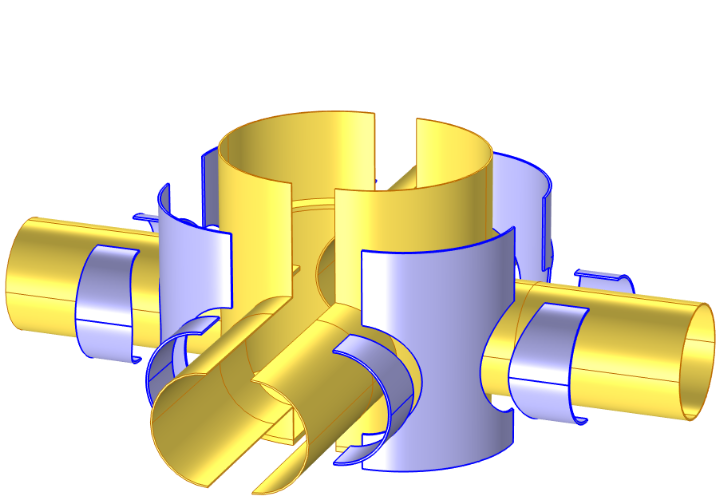}
    \caption{\footnotesize Schematic drawing of the lower part of a TM tower (yellow) and a possible ferromagnetic shielding (blue).}
    \label{fig:rivestimento mu-metal}
\end{minipage}
\end{figure}

The test masses (TMs) are the most sensitive components of the interferometer and must be shielded from any form of noise.\\[0.1cm]
In the Virgo interferometer, the TMs are equipped with four magnets arranged in an antisymmetric configuration (Fig.~\ref{fig:TM_magnets}). These magnets enable precise mirror control through four coils positioned on the payload behind the mirror.\\[0.1cm]
However, if the magnets on TM interact with magnetic fields other than the intended control fields, they inevitably generate noise. To prevent this, it is crucial to shield this region of the interferometer from external magnetic fields. A possible solution consists in enclosing the lower part of the tower and the arms near the intersection area with ferromagnetic shielding. A possible configuration is illustrated in Fig.~\ref{fig:rivestimento mu-metal}, where a mu-metal layer is applied to the lower part of the tower, covering a height of 2.375~m, and extends 0.585~m along each of the four arms.

%%%%%%%%%%%%%%%%%%%%%%%%%%%%%%%%%%%%%%%%%%%%

\section{Results}\label{Results}
\begin{figure}
    \begin{minipage}[c]{0.44\textwidth}
        \centering
        \includegraphics[width=0.9\linewidth]{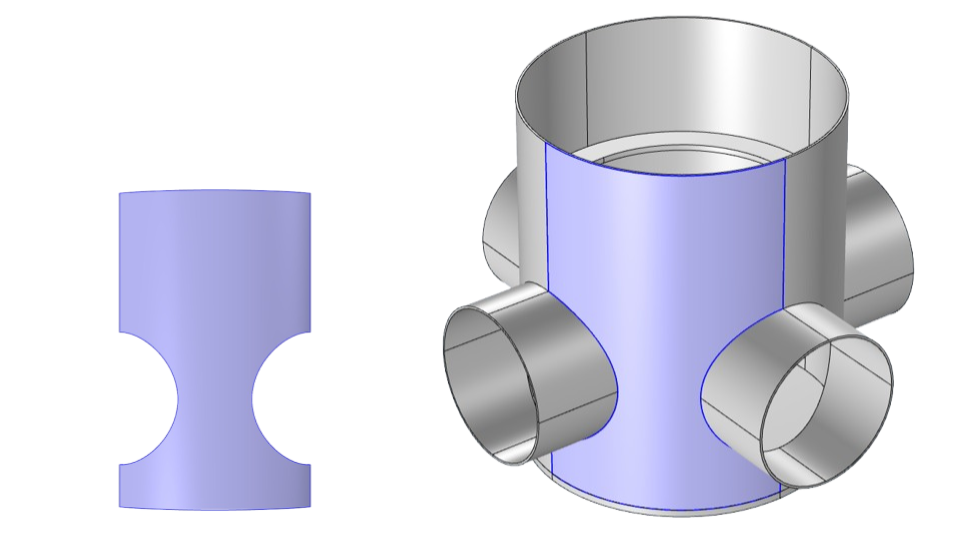}
        \caption{\footnotesize A simplified geometry of the lower part of a Virgo tower is depicted in gray, while one of the ferromagnetic shields employed in the simulation is highlighted in blue.}
        \label{fig:senza viewports}
    \end{minipage}
    \hfill
    \begin{minipage}[c]{0.44\textwidth}
        \centering
        \includegraphics[width=0.9\linewidth]{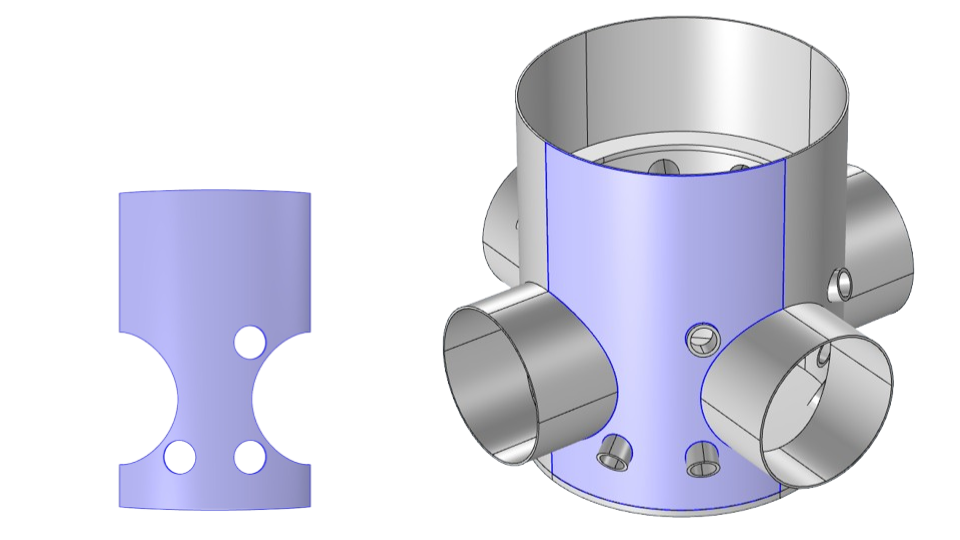}
        \caption{\footnotesize A more realistic representation of the tower, including the addition of the viewports, is depicted in gray, while one of the ferromagnetic shields employed in the simulation is highlighted in blue.}
        \label{fig:con viewports}
    \end{minipage}
\end{figure}

\begin{figure}[h!]
    \centering
    \includegraphics[width=0.45\linewidth]{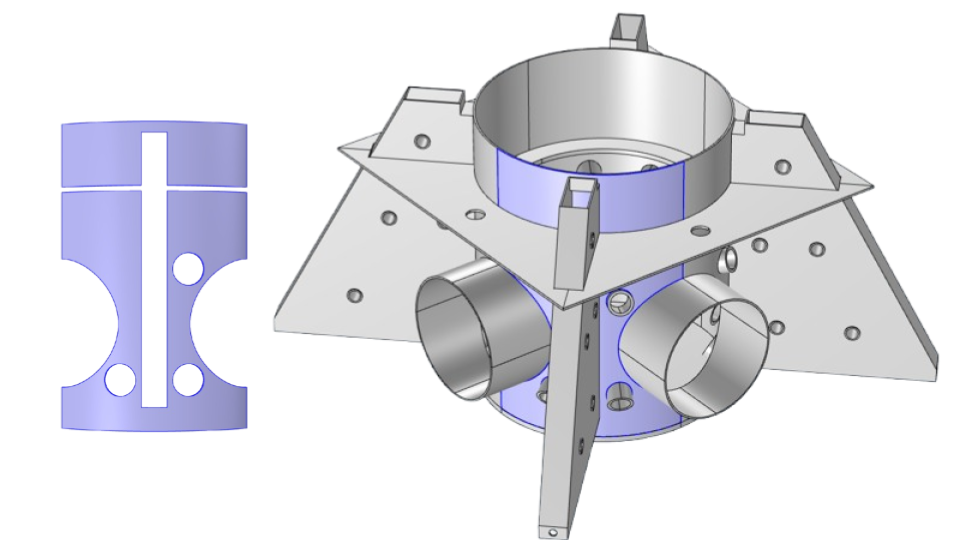}
    \caption{\footnotesize A more realistic representation of the tower, including the viewports and the mechanical supports, is depicted in gray, while one of the ferromagnetic shields employed in the simulation is highlighted in blue.}
    \label{fig:torre reale}
\end{figure}

{\footnotesize T}{\scriptsize ABLE}~{\footnotesize\ref{table:SF_torre_piena}} shows the shielding factor obtained from simulations using the relative magnetic permeability value reported by \cite{Arpaia} at various frequencies under an external magnetic field of few $nT$. The underlying assumption is that for the field strengths of interest in our study, i.e. from $pT$ to $nT$~\cite{Schumann_Virgo},~\cite{Magnetic_Noise_Virgo},~\cite{ET-0069A-22}, the permeability remains approximately constant, i.e. the material maintains a linear response below the nanotesla threshold.\\[0.1cm]
Since no definitive designs exist yet for the Einstein Telescope towers, the geometric design considered is that of Virgo: the tower has a radius of 1~m, while each arm has a radius of 0.5~m (Fig.~\ref{fig:senza viewports}).\\[0.1cm]
Simulations were conducted using COMSOL Multiphysics, a finite element software. The Shielding Factor is defined as the ratio of the incoming flux, measured before and after the implementation of the mitigation system, within a sphere of 0.5~m radius centered at the intersection point between the tower and its arms.
\begin{table}[h!]
    \centering
    \caption{\footnotesize Relative magnetic permeability of mu-metal at various frequencies under an external magnetic field on the order of nanotesla, and the corresponding shielding factor achieved using a 1~mm/2~mm thick mu-metal layer to attenuate a magnetic field directed along one arm of the interferometer. The simulated geometry is shown in Fig.~\ref{fig:senza viewports}.}
    \label{table:SF_torre_piena}
    \begin{tabularx}{\textwidth}{>{\centering\arraybackslash}p{0.23\textwidth}|>{\centering\arraybackslash}p{0.28\textwidth}>{\centering\arraybackslash}p{0.15\textwidth}>{\centering\arraybackslash}p{0.15\textwidth}}
    \hline
    \rule{0pt}{15pt} \textbf{Frequency} & \textbf{Relative} & \multicolumn{2}{c}{\textbf{Shielding Factor}} \\ 
    \rule{0pt}{12pt} \textbf{[Hz]} & \textbf{Permeability} & width=1~mm & width=2~mm \\ [0.5ex] 
    \hline\hline
    \rule{0pt}{12pt} $0$ & $57\,000$ & $10.40$ & $14.47$ \\
    \rule{0pt}{12pt} $10$ & $48\,000$ & $9.43$ & $13.47$ \\
    \rule{0pt}{12pt} $100$ & $40\,000$ & $8.45$ & $12.39$ \\
    \rule{0pt}{12pt} $1000$ & $15\,000$ & $4.41$ & $7.04$ \\[0.1cm]
 \hline
    \end{tabularx}
\end{table}\\
{\footnotesize T}{\scriptsize ABLE}~{\footnotesize\ref{table:SF_torre_bucata}} presents the same quantities calculated on a more realistic model of the tower, which includes the viewports present in the Virgo towers (Fig.~\ref{fig:con viewports}). At these positions, the application of a mu-metal layer is not feasible.\\[0.1cm]
\begin{table}[h!]
    \centering
    \caption{\footnotesize Relative magnetic permeability of mu-metal at various frequencies under an external magnetic field on the order of nanotesla, and the corresponding shielding factor achieved using a 1~mm/2~mm thick mu-metal layer to attenuate a magnetic field directed along one arm of the interferometer. The simulated geometry is shown in Fig.~\ref{fig:con viewports}.}
    \label{table:SF_torre_bucata}
    \begin{tabularx}{\textwidth}{>{\centering\arraybackslash}p{0.23\textwidth}|>{\centering\arraybackslash}p{0.28\textwidth}>{\centering\arraybackslash}p{0.15\textwidth}>{\centering\arraybackslash}p{0.15\textwidth}}
    \hline
    \rule{0pt}{15pt} \textbf{Frequency} & \textbf{Relative} & \multicolumn{2}{c}{\textbf{Shielding Factor}} \\ 
    \rule{0pt}{12pt} \textbf{[Hz]} & \textbf{Permeability} & width=1~mm & width=2~mm \\ [0.5ex] 
    \hline\hline
    \rule{0pt}{12pt} $0$ & $57\,000$ & $9.89$ & $13.85$ \\
    \rule{0pt}{12pt} $10$ & $48\,000$ & $8.95$ & $12.87$ \\
    \rule{0pt}{12pt} $100$ & $40\,000$ & $8.02$ & $11.82$ \\
    \rule{0pt}{12pt} $1000$ & $15\,000$ & $4.19$ & $6.67$ \\[0.1cm]
 \hline
    \end{tabularx}
\end{table}\\
\\ 
Finally, {\footnotesize T}{\scriptsize ABLE}~{\footnotesize\ref{table:SF_torre_molto_bucata}} presents the same quantities for an extremely realistic model of the tower, which incorporates the viewports and mechanical supports found in the Virgo towers (Fig.~\ref{fig:torre reale}). For these areas, no ferromagnetic shielding is applied.\\[0.1cm]
\begin{table}[h!]
    \centering
     \caption{\footnotesize Relative magnetic permeability of mu-metal at various frequencies under an external magnetic field on the order of nanotesla, and the corresponding shielding factor achieved using a 1~mm/2~mm thick mu-metal layer to attenuate a magnetic field directed along one arm of the interferometer. The simulated geometry is shown in Fig.~\ref{fig:torre reale}.}
    \label{table:SF_torre_molto_bucata}
    \begin{tabularx}{\textwidth}{>{\centering\arraybackslash}p{0.23\textwidth}|>{\centering\arraybackslash}p{0.28\textwidth}>{\centering\arraybackslash}p{0.15\textwidth}>{\centering\arraybackslash}p{0.15\textwidth}}
    \hline
    \rule{0pt}{15pt} \textbf{Frequency} & \textbf{Relative} & \multicolumn{2}{c}{\textbf{Shielding Factor}} \\ 
    \rule{0pt}{12pt} \textbf{[Hz]} & \textbf{Permeability} & width=1~mm & width=2~mm \\ [0.5ex] 
    \hline\hline
    \rule{0pt}{12pt} $0$ & $57\,000$ & $4.07$ & $6.10$ \\
    \rule{0pt}{12pt} $10$ & $48\,000$ & $3.69$ & $5.51$ \\
    \rule{0pt}{12pt} $100$ & $40\,000$ & $3.34$ & $4.95$ \\
    \rule{0pt}{12pt} $1000$ & $15\,000$ & $2.10$ & $2.88$ \\[0.1cm]
 \hline
    \end{tabularx}
\end{table}\\
%%%%%%%%%%%%%%%%%%%%%%%%%%%%%%%%%%%%%%%%%%%%

\section{Conclusion}\label{Conclusion}
The relative magnetic permeability values used ({\footnotesize T}{\scriptsize ABLE}~{\footnotesize\ref{table:SF_torre_piena}}, {\footnotesize T}{\scriptsize ABLE}~{\footnotesize\ref{table:SF_torre_bucata}}, {\footnotesize T}{\scriptsize ABLE}~{\footnotesize\ref{table:SF_torre_molto_bucata}}) refer to samples that have undergone an annealing process. Consequently, to preserve their optimized magnetic properties, mu-metal components must be manufactured to their final geometry prior to annealing. Any subsequent mechanical processing would introduce internal stresses and defects, significantly degrading the material’s magnetic performance.\\[0.4cm]
The shielding factor in the second configuration (Fig.~\ref{fig:con viewports}) exhibits a slight reduction compared to the first (Fig.~\ref{fig:senza viewports}); however, this difference is not significant. This outcome aligns with expectations, considering that the openings in the ferromagnetic shield at the viewports are substantially smaller than those at the beam pipe interfaces: the viewport radius is 0.08~m, whereas the arm openings have a radius of 0.5~m. Consequently, the magnetic flux penetrating through the viewports is considerably less than that entering through the larger arm openings.\\[0.1cm]
In contrast, the third configuration (Fig.~\ref{fig:torre reale}) exhibits a substantial decrease in the shielding factor, attributable to the removal of a significantly larger portion of the shielding surface.\\[0.4cm]
The shielding factor obtained using the most realistic model ({\footnotesize T}{\scriptsize ABLE}~{\footnotesize\ref{table:SF_torre_molto_bucata}}) with a 2~mm thick layer is a factor of 5 at low frequencies ($10-100~Hz$) and a factor of 6 at very low frequencies ($<10~Hz$). These values suggest that, assuming the magnetic noise is mainly due to coupling with the test masses, the implementation of such a mitigation system would be sufficient to adequately reduce natural magnetic noise - especially considering that the situation could be further improved by employing multilayer shielding.\\[0.1cm]
In contrast, these values are insufficient to address the issue of self-inflicted noise. To resolve this problem, as outlined in Section \ref{Magnetic Noise}, a triple strategy must be adopted, involving the shielding of the magnetic noise source itself and the reduction of the coupling mechanism.
%\acknowledgments
%The author acknowledge XXX, YYY. \cite{SiteSelection}, \cite{ref:pul}

%%%%%%%%%%%%%%%%%%%%%%%%%%%%%%%%%%%%%%%%%%%%
% bibliografia

\end{document}